\documentclass[10pt, a4paper, journal, draftclsnofoot, onecolumn]{IEEEtran}

\usepackage[pdftex]{graphicx}
\usepackage{bm,afterpage,amsmath,hyperref,placeins,flafter}
\usepackage[numbers,sort]{natbib}

\graphicspath{{figures/}}

\begin{document}

\title{The Relation Between Global Migration and Trade Networks}

%
%
%

\author{Paolo Sgrignoli, Rodolfo Metulini, Stefano Schiavo, Massimo Riccaboni
\IEEEcompsocitemizethanks{\IEEEcompsocthanksitem Paolo Sgrignoli, Rodolfo Metulini and Massimo Riccaboni, IMT Advanced Studies Lucca - Laboratory of Innovation Management and Economics; Stefano Schiavo, School of International Studies and Department of Economics and Management, University of Trento \protect}
\thanks{Manuscript submitted Oct. 23 2013}}

\maketitle

\begin{abstract}
\noindent In this paper we develop a methodology to analyze and compare multiple global networks. We focus our analysis on the relation between human migration and trade. First, we identify the subset of products for which the presence of a community of migrants significantly increases trade intensity. To assure comparability across networks, we apply a hypergeometric filter to identify links for which migration and trade intensity are both significantly higher than expected. Next we develop an econometric methodology, inspired by spatial econometrics, to measure the effect of migration on international trade while controlling for network interdependencies. Overall, we find that migration significantly boosts trade across sectors and we are able to identify product categories for which this effect is particularly strong.

\end{abstract}


%
\IEEEpeerreviewmaketitle

\section{Introduction}

Since the mid 1990s a growing body of research has investigated the relation between global networks of international trade, information flows and migrations. Whereas traditional trade theory (e.g. the standard Heckscher-Ohlin model) suggests that the movement of goods across borders can provide a substitute for the movement of production factors, the bottom line of this branch of research is that the two actually complement each other. This appears to hold for different countries  \cite[for the US, Canada and Spain respectively, see][]{gould1994immigrant,head1998immigration,peri2010trade} and has recently been confirmed by a meta-analysis covering 48 different studies \cite{gen2011impact}.

Most of the literature referred to above shares a common empirical strategy, based on the estimation of a log-linear gravity model where bilateral trade flows are regressed over standard explanatory variables (economic mass and distance), the stock of immigrants from specific partner countries and other controls aiming at capturing various types of trade costs (common language, colonial relationships and the like). The most influential contributor to the field is probably James Rauch who, in a series of papers with a variety of co-authors, looks at the role of ethnic immigrants networks in facilitating trade. The core of the argument is that formal and informal links among co-ethnic migrants in other countries and at home, the "network" in Rauch's terminology, facilitate trade by providing potential trading partners with easier access to valuable, i.e. qualified, information. The pro-trade effect thus stems from the reduction of the trade barriers and search costs associated with market transactions. Since these costs are likely to be larger for international trade due to distance, language and cultural differences, legal provisions and the like, "networks" end up being especially relevant in facilitating cross-border transactions. Indeed, the central result in Rauch and Trindade \cite{rauch2002ethnic} is that the positive effect of migration on trade is larger for "differentiated goods", i.e. those items that are not homogeneous and are not traded in organized exchanges therefore rendering that knowledge about counterpart reputation particularly valuable.

Similar results have been replicated by a number of subsequent authors. Peri and Requena-Silvente \cite{peri2010trade}, for instance, analyze the Spanish case and find that doubling the number of immigrants from a given country increases export to the same destination by 10 percent. This effect is higher for firms selling differentiated products and for more distant countries (geographically or culturally). All of these elements are consistent with the notion that networks (in this case the presence of a large community of expatriates and their connections with co-nationals at home and abroad) lower the hurdle in terms of economic interactions, providing better access to information and trade opportunities and reducing the fixed costs associated with entry into a foreign market.

Easier access to information via co-ethnic migrants does not only increase export from the recipient to the home countries, but also facilitates "chain migration": it is typical for communities with a significant presence of migrants to attract more migrants from the same community. This feature is consistent with the preferential attachment mechanism that accurately describes the evolution of many real-world networks (from airline traffic to the World Wide Web, from social ties to financial networks) and makes migration interesting in terms of complex network analysis. Yet, not much has been written on the subject: Slater \cite{slater2008hubs} studies clustering in US internal migration, whereas Simini et al. \cite{simini2012universal} presents a stochastic radiation model that can be used to predict international migration patterns.

More recently, Fagiolo and Mastrolillo \cite{fama13pre} study the topology of the migration network, and its evolution over the period 1960-2000. They find that the network (where links between two countries A and B are given by the stock of migrants originated in country A and living in country B in a given year) is disassortative and highly clustered, and displays a small-world binary pattern. Furthermore, they show that the structural properties of the network are mainly driven by socioeconomic, geographical, and political factors. In a second paper \cite{fama13arXiv} they look at the interplay between trade and migration, finding that the two networks are strongly correlated, and country-pairs that are more central in the migration network tend to trade more.

In this paper we aim at filling this gap by analyzing the relationship between migration and trade through the lens of complex network analysis and spatial econometrics. To do so, we first make the networks comparable, using a hypergeometric benchmark to judge if the link intensity between two countries is significantly higher than expected. Next, we run a set of econometric regressions, checking for network interdependencies: by means of spatial econometric techniques, we check for the presence of network (auto)correlations between trade and migration, disentangling the analysis in different parts, and using the previously defined network matrices as weights in spatial autoregressive models. Our goal is to identify the subset of products whose export/import is the most highly related to the presence of a community of migrants.

This paper is structured as follows. In Section \ref{data} we describe migration and trade data, as well as our methodological approach. In Section \ref{network} we analyze and compare the migration and trade networks by distinguishing different product types. In Section \ref{econ_app} we presents the econometric analysis and discuss the findings. Finally, Section \ref{concl} contains discussions of our main results, further research directions and conclusions.

\section{Data \& Methodology}\label{data}

\subsection{Migration and trade data}
In this paper we use data regarding migrants from The World Bank's Global Bilateral Migration dataset \cite{ozden2011earth}: it is composed by decennial matrices of bilateral migrant stocks spanning 1960-2000 (5 census rounds), disaggregated by gender and based primarily on the foreign-born definition of migrants. It is the first and only comprehensive picture of bilateral global migration over the second half of the 20th century, taking into account a total of 232 countries. The data reveal that the global migrant stock increased from 92 million in 1960 to 165 million in 2000. Quantitatively, migration between developing countries dominates, constituting half of all international migration in 2000, whereas flows from developing to developed countries represent the fastest growing component of international migration in both absolute and relative terms. As for international trade, we use the dataset that was cleaned and made compatible through a National Bureau of Economic Research (NBER) project led by R. Feenstra \cite{NBERw11040}, disaggregated according to the Standardized International Trade Code at the four-digit level (SITC-4). It provides for each country the value (expressed in thousands of US dollars) exported to all other countries, for 775 product classes. In our analysis, we focus on the years 1970 and 2000, although choosing different years does not alter the results.

In addition, to improve our analysis, we look at the SITC product code traded between each country pair and perform a decomposition of goods following the Rauch classification \cite{rauch2002ethnic}. We thus distinguish between \textit{homogeneous} and \textit{differentiated} goods: the former are those with a reference price, whether it be a result of organized exchanges or simply of price quotations in a specialized journal; while the latter are without such a reference price and can be thought of as 'branded' commodities.
An important distinction between these two categories is that homogeneous product prices can be quoted without mentioning the name of the manufacturer, like many chemical compounds for example, letting specialized traders stay informed regarding their prices around the globe and perform international commodity arbitrage by matching distant buyers and sellers much like traders in an organized exchange. In other words for this category of products it is immediately apparent if the differential between the two countries' markets is large enough to cover customs and transportation costs.
On the contrary, differentiated products are such that their commodity categories are usually not well defined (e.g. footwear); they need to be disaggregated into various sub-types, a process that leads to the limit where each category contains only one supplier. These products are, in this respect, 'branded' or differentiated.
From this definition one would expect for international trade to be more heavily influenced by migrant networks for this second class of products, as buyers and sellers need to be matched in the product characteristics space. This is indeed the result found by Rauch \cite{rauch2002ethnic} and one of the aspects we test in this paper.

Our aim is to use the two datasets together. In this respect, we retain only the countries present in both of them to enhance comparability. Doing this we end up with a set of 146 nodes, that populate both the trade and migration networks.

\subsection{Methodology}\label{method}
The different nature of the data, both with regards to type and measurement unit, make a direct comparison of the trade and migration networks unreliable. As such what is needed is a tool to make the two datasets comparable. To address this issue, we use a stochastic benchmark for normalization purposes, as recently used in \cite{riccaboni2013global} and previously introduced in \cite{tumminello2011statistically}. This method has been used to identify statistically significant portions of data in fields ranging from genetics to network theory \cite{tavazoie1999systematic,wuchty2006stable}.

For two countries, $A$ and $B$, let $N_A$ be the value of goods exported by country $A$ and $N_B$ the value of goods imported by country $B$. The total value of traded goods is $N_k$ and the observed value of goods exported from $A$ to $B$ is $N_{AB}$. Under the null hypothesis of random co-occurrence, that is to say, customers in country $B$ who are indifferent to the nationality of the exporter, the probability of observing $X$ thousands US dollars of goods traded is given by the hypergeometric distribution
\begin{equation}
H(X|N_k,N_A,N_B) = \frac{{N_A \choose X} {{N_k-N_A} \choose {N_B-X}}}{{N_k \choose N_B}} \ ,
\end{equation}
and we can associate a \textit{p}-value with the observed $N_{AB}$ as
\begin{equation}
p(N_{AB}) = 1 - \sum_{X=0}^{N_{AB}-1} H(X|N_k,N_A,N_B) \ .
\end{equation}
Note that the described null hypothesis directly takes into account the heterogeneity of countries with respect to the total value of goods traded (row and column totals). For each pair of countries, we separately evaluate the \textit{p}-value for each trade relationship and then use a cutoff to select only those links that represent a significant departure from the hypergeometric benchmark ($p<.01$). The resulting matrices are then dichotomized\footnote{The hypergeometric multi-urn benchmark is equivalent to the Monte Carlo degree-preserving network rewiring procedure \cite{maslov2002specificity}}. Thanks to our stochastic approach we can treat a weighted network as a random graph in which any link has a probability of occurrence. We use this method to filter both trade and migration networks, to make them comparable.

In the next Section we study the structural properties of trade and migration networks. We use then networks as weight matrices in a spatial gravity model of trade flows (Section \ref{econ_app}).

\section{Network analysis}\label{network}

We first describe and compare some basic topological properties of the two networks separately (Section \ref{top_prop}); we then to analyze the interconnections and correlations among their links (Section \ref{links_corr}).

\subsection{Topological properties}\label{top_prop}
In this section we introduce and analyze some topological indicators of the \textit{migration} and \textit{trade} (both total and separated in differentiated and homogeneous products) networks for the year 2000.

\begin{figure}[!htb]
\centering
\includegraphics[width=\columnwidth]{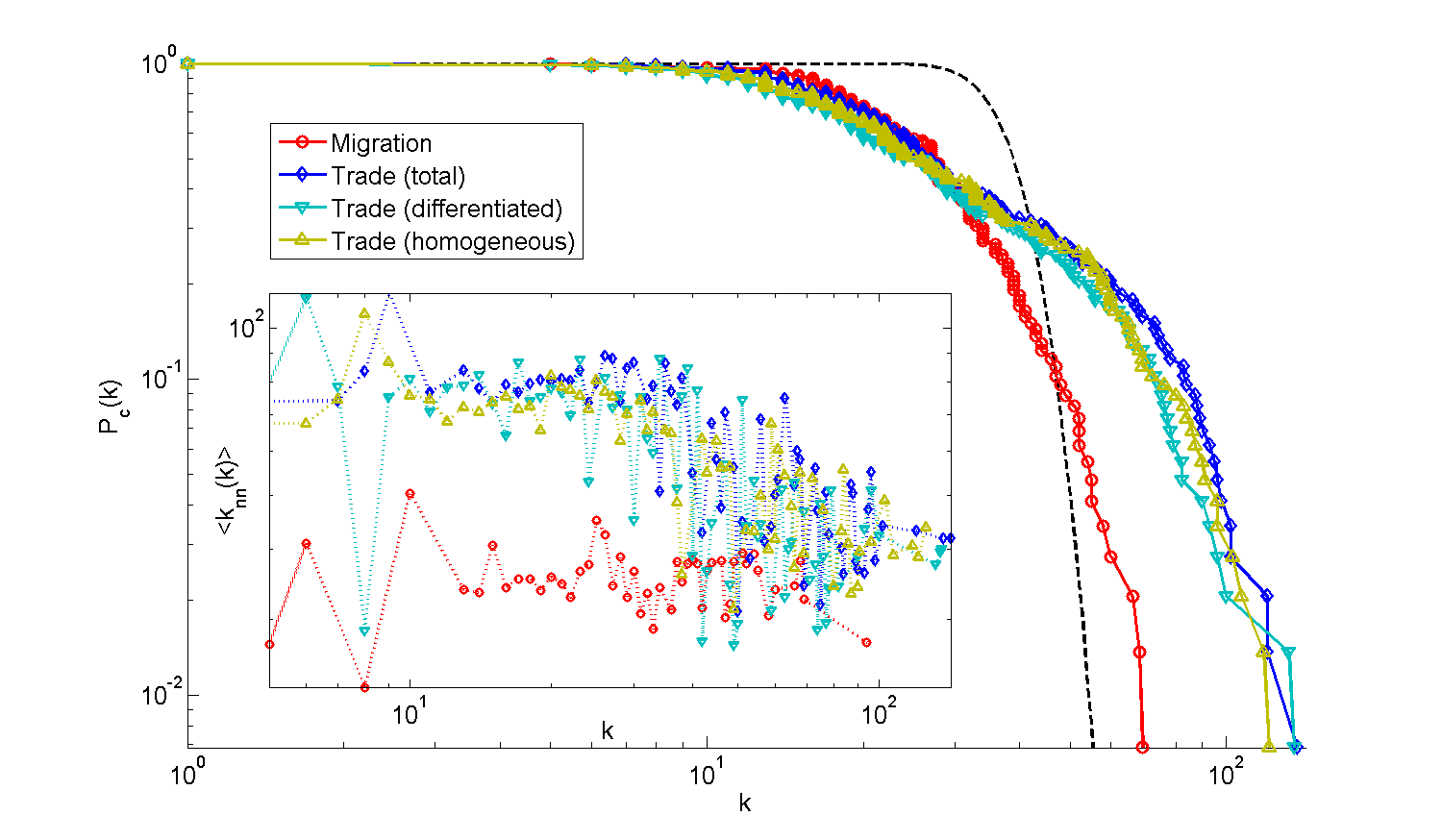}
\caption{Main: cumulative degree distribution $P_c(k)$ for the trade and migration networks. The dashed line is the same degree distribution for a random graph with the same average degree. Inset: average nearest neighbors' degree as a function of the total degree of the vertex.}
\label{fig:annd}
\end{figure}

One of the most important topological properties of a network is the degree distribution, $P(k)$. This quantity measures the probability of a randomly chosen vertex to have $k$ neighbors. Figure \ref{fig:annd} (main) shows the undirected cumulative distribution of trade and migration networks, defined as $P_c(k)\equiv \sum_{k'>k}P(k')$. In all cases the cumulative distributions show a flat approach from the origin, similar to the Erd\"{o}s-R\'{e}nyi (ER) network; but then, at about $k=10$, the distributions follow a power law decay, with a strong deviation from the exponential tail predicted by the classical random graph theory. This is especially true for the trade networks' cumulative distribution, whereas the migration case has a less evident power law behavior and deviation from the ER case.

Table \ref{tab:top} presents a selection of well-known statistics for these networks. We can observe that the trade networks share, in general, very similar properties, although they are always slightly different for the migration case. In particular, while the migration network has a lower connectivity, as indicated by the nodes' average degree and density, its links are more reciprocal, meaning that a bilateral bond is more common than in commercial trades.

\begin{table}[!ht]
\centering
\caption{Year 2000. D = Differentiated; H = Homogeneous.}
\begin{tabular}{lcccc}
\hline
\rule{0pt}{2ex} & Migration & Trade & Trade (D) & Trade (H)\\
\hline
\rule{0pt}{3ex}Average degree & 29.45 & 38.47 & 34.30 & 36.19\\
Density (\%) & 15.1 & 20.4 & 18.5 & 19.3\\
Corr. coeff. & 0.45 & 0.79 & 0.70 & 0.78\\
Reciprocity (\%)& 34.5 & 30.2 & 28.2 & 29.1\\
Average cluster coeff. & 0.099 & 0.088 & 0.073 & 0.089\\
Assortativity & -0.009 & -0.428 & -0.278 & -0.368\\
\hline
\end{tabular}
\label{tab:top}
\end{table}

Another important aspect is the hierarchical structure of the networks, which is usually analyzed by means of the clustering coefficient and degree-degree correlation. The \textit{clustering coefficient} of vertex $i$ is defined as $c \equiv 2n_i/k_i(k_i-1)$, where $n_i$ is the number of neighbors of $i$ that are interconnected. The values of the clustering coefficient for our networks are $\simeq 0.1$, much lower than in other analyses of the world trade network \cite[See for example][]{serrano2003topology}, indicating that the hypergeometric filtering we used in our work to dilute it resulted in the loss of the clustered structure of the network. Hierarchy is also reflected in the degree-degree correlation through the conditional probability $P(k|k')$, i.e. the probability that a vertex of degree $k'$ is linked to a vertex of degree $k$. This function is difficult to measure, due to statistical fluctuations, and it is usually substituted by the \textit{average nearest neighbors degree (ANND)}, defined as $\left\langle k_{nn}(k)\right\rangle =\sum_{k'}k'P(k'|k)$ \cite{pastor2001dynamical}. For independent networks this quantity would result independent of $k$. Figure \ref{fig:annd} (inset) reports the ANND for the migration and trade networks, showing a dependency on the vertex's degree and indicating that in all the networks highly connected vertices tend to connect to poorly connected vertices (which is usually called \textit{dissortative} behavior). It is clear how this phenomenon is much more pronounced in trade than in migration, as can also be seen by the assortativity coefficient\footnote{Also known as \textit{Pearson's degree correlation}} in Table \ref{tab:top}.

\subsection{The Relation between migration and trade}\label{links_corr}
A first approach to analyzing the interplay between international trade and human migration is to observe whether, for a pair of countries, a strong bond in one aspect is related to a strong tie in anther. To do this we make use of the Jaccard index \cite{jacc1901} that, given two sets of events, is defined as $J=\left| A\cap B\right| /\left| A\cup B\right|$, therefore representing the ratio between the number of events shared between the two sets, over the number of events in at least one of them. In our case the events will represent that the link between two countries is significant\footnote{Significant in the sense of Section \ref{method}, so to say with the \textit{p}-value below the threshold we set at $0.01$.} both in the trade and the migration networks.

\begin{figure}[!htb]
\centering
\includegraphics[width=\columnwidth]{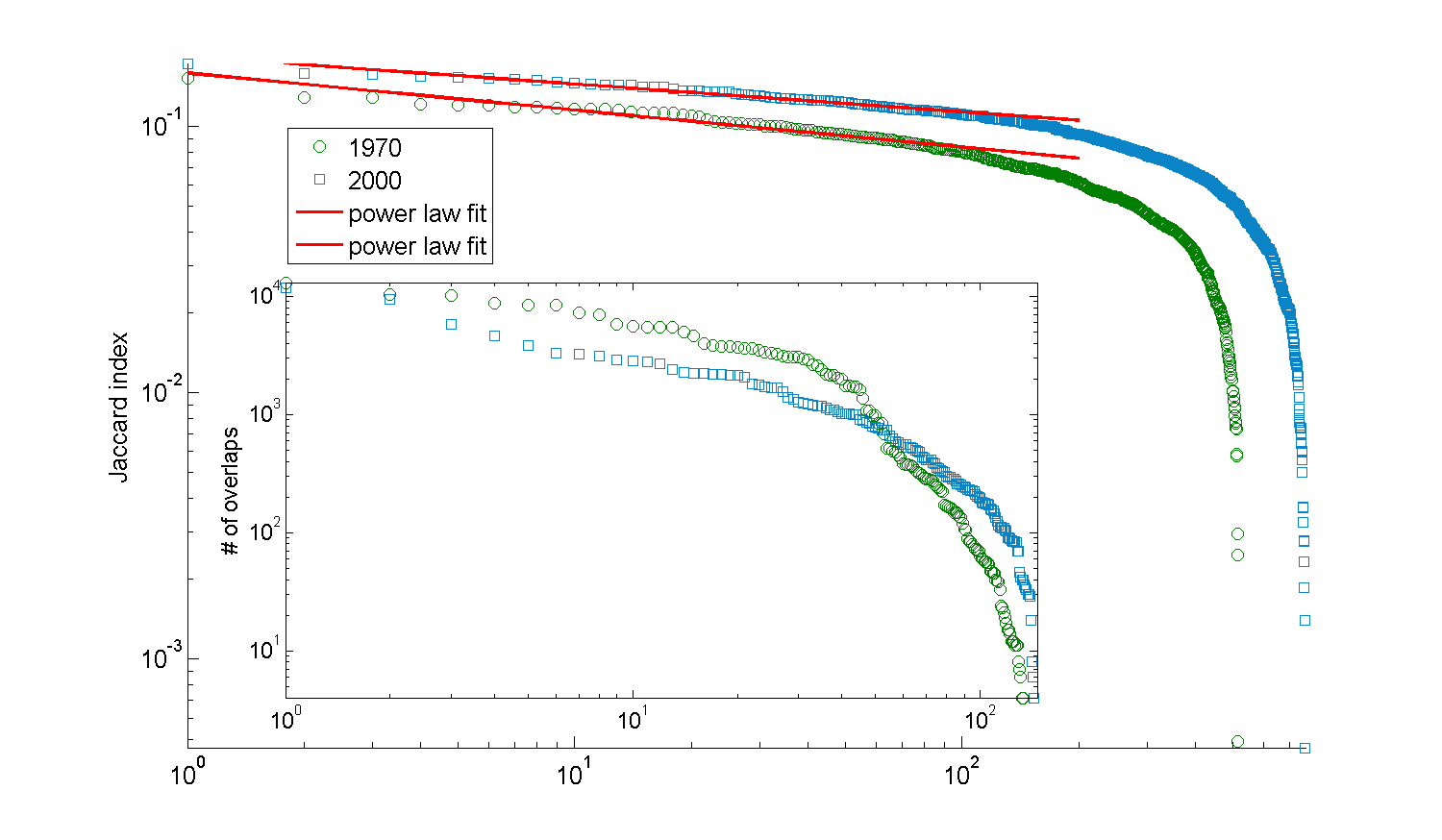}
\caption{Main figure: Distribution of the Jaccard index relative to the overlapping of single products with the migration networks for years 1970 and 2000. Red lines are power law fits. Inset: Distribution of the number of overlaps between the trade and migration networks for individual countries.}
\label{fig:total}
\end{figure}

In Figure \ref{fig:total} (main) we can see the distribution of the Jaccard index ($J$) for all the individual products in the years 1970 and 2000: both present a clear power law behavior and show how for products for the year 2000, the overlap with migration is generally higher. In the inset we plot the total number of overlaps in the networks for individual countries: this too presents a power-law behavior, although the two distributions cross at some point, indicating that countries with a stronger relation between trade and migration had a higher network effects in 1970, while countries with a weaker bond had a higher one in 2000.

In Tables \ref{tab:countries_list} and \ref{tab:products_list} we also list the top and bottom countries and products, respectively, that appear in the distributions.

\begin{figure}[!ht]
\centering
\includegraphics[width=\columnwidth]{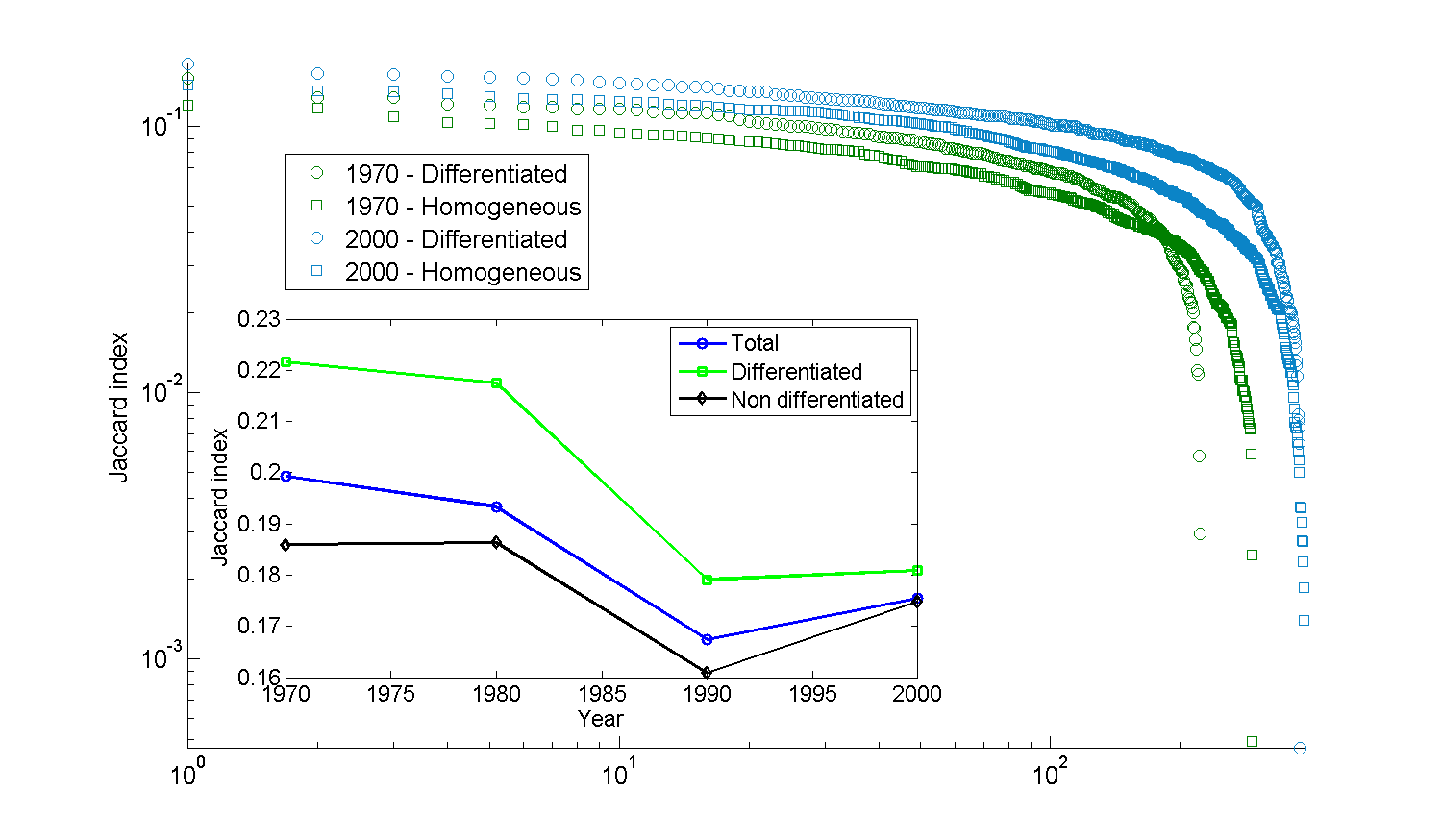}
\caption{Main figure: Distribution of the Jaccard index relative to the overlapping of single products with the migration networks, shown separately for \textit{differentiated} and \textit{homogeneous} products, for years 1970 and 2000. Inset: Jaccard indexes relative to the overlap of the entire trade and migration networks for different years. A distinction is made between all the products and just differentiated or homogeneous ones.}
\label{fig:diff}
\end{figure}

To verify the validity of the Rauch classification \cite{rauch2002ethnic} we repeated the same analysis separating \textit{differentiated} and \textit{homogeneous} goods: in the main panel of Fig. \ref{fig:diff} we plot the distribution of the Jaccard index for individual products while in the inset we show the Jaccard index for the entire networks in the years 1970-2000. As one can observe, in this first analysis the Rauch classification seems to be verified by our results, with values for the differentiated products generally higher for both individual SITC codes and the whole networks.

To further investigate this point, in Section \ref{econ_app} we compare the results for the classification proposed by Rauch and the that emerges from the overlapping of the trade and migration networks (see Table \ref{tab:products_list}). In particular, we take the \emph{top-N} products\footnote{\emph{N} indicates here the number of differentiated products in each year.} from our ranking and compare them with Rauch's differentiated ones, and do the same for the bottom ranked with homogeneous commodities. We call these two new categories \emph{overlapping} and \emph{non-overlapping}\footnote{In our new classification, for the year 1970 the overlapping (non-overlapping) products are composed by $54\%-46\%$ ($27\%-73\%$) of Rauch's differentiated-homogeneous products; while for the year 2000 the figure changes to $62\%-38\%$ ($37\%-63\%$)}.

\begin{table}[!ht]
\centering
\caption{A list of the countries that best and worst overlap between trade and migration networks, based on the Jaccard index, year 2000.}
\begin{tabular}{cl}
\hline
\rule{0pt}{2ex} \textbf{\#} & \textbf{Country}\\
\hline
\rule{0pt}{3ex} 1 & Germany\\
2 & United Kingdom\\
3 & Italy\\
4 & Spain\\
5 & Netherlands\\
$\cdots$ & $\cdots$\\
11 & China\\
$\cdots$ & $\cdots$\\
142 & Sierra Leone\\
143 & Somalia\\
144 & Saint Pierre and Miquelon\\
145 & Tanzania, United Republic of\\
146 & Uganda\\
\hline
\end{tabular}
\label{tab:countries_list}
\end{table}

\begin{table}[!h]
\centering
\caption{A list of the products (SITC-4) with the most and least overlap with the migration network, based on the Jaccard index, for year 2000. D = Differentiated; H = Homogeneous.}
\begin{tabular}{clc}
\hline
\rule{0pt}{2ex} \textbf{\#} & \textbf{Product} & \textbf{Type}\\
\hline
\rule{0pt}{3ex} 1 	& Miscellaneous articles of plastic	& D \\
2 	& Insulated electric wire. cable. bars. etc	& D \\
3 	& Plastic packing containers. lids. stoppers and other closures	& D \\
4 	& Switches. relays. fuses...; switchboards and control panels... & D \\
5 	& Edible products and preparations. nes	& D \\
6 	& Chemical products and preparations. nes & D \\
7 	& Other furniture and parts thereof. nes & D \\
8 	& Machinery for specialized industries and parts thereof. nes & D \\
9 	& Other polymerization and copolymerization products & D \\
$\cdots$ & $\cdots$ & $\cdots$\\
762 & Copra & H \\
763 & Palm nuts and kernels & H \\
764 & Manila hemp. raw or processed but not spun. ... & H\\
765 & Uranium depleted in U235. thorium. and alloys. nes; ... & H \\
766 & Ores and concentrates of uranium and thorium & H \\
767 & Castor oil seeds & D \\
768 & Coal gas. water gas and similar gases & H \\
769 & Wood-based panels. nes & D \\
770 & Knitted or crocheted fabrics. elastic or rubberized & D \\
\hline
\end{tabular}
\label{tab:products_list}
\end{table}

\section{Econometric approach}\label{econ_app}

Gravity models have often been used to explain \emph{Origin-Destination} (OD) flows that arise in fields such as trade, transportation and migration. The term \emph{spatial interaction models} has been used in the literature to label models that focused on flow between origins and destinations \cite{lee2005spatial}. A great deal of literature on theoretical foundations for this model exists \cite{anderson1979theoretical,van2003gravity}, as well as literature focused on empirical studies \cite{frankel2000estimate,egger2000note,anderson1979theoretical,van2003gravity}. These models rely on a function of the distance between an origin and destination in addition to explanatory variables pertaining to characteristics of both origin and destination regions. Spatial interaction models assume that using distance as a variable will eradicate the spatial dependence among the OD pairs. However, the notion that using of distance functions in conventional interaction models effectively captures spatial dependence in interregional flows has long been challenged.

It has been recently fully recognized that a spatial interaction effect exists \cite{bang2006regional,kelejian2012neighborhood,baltagi2007estimating,hall2008spatial}, essentially due to the spatial spillover and the third country effect. This motivates the introduction of the spatial autoregressive components in the so-called \emph{spatial gravity model of trade}. The spatial autoregressive component makes use of an $n\times n$ weight matrix in order to define the set of the spatial neighbors. The correct specification of the weight matrix is still debated. It is, however, reputed that the correct specification of this matrix, conditioned to the topic and to the objectives of the analysis, exists. One of the most frequently used weight matrices is based on spatial contiguity: it contains 1 if the pair shares geographical borders, 0 otherwise.
$$
\mbox{W: } w_{i,j} = 1 \mbox{ if $i$ is neighbor of $j$, } w_{i,j}= 0 \mbox{ otherwise.}
$$
Other formulations are also proposed based on an inverse distance approach, as well as on the similarity of technologies in the pair or regions. It is important to underline that the neighboring regions of OD flows include neighbors of the origin location, neighbors of the destination location, and perhaps a link between neighbors of the origin and neighbors of the destination regions. To account for this, the matrix dimension that suits in this case is $n^2 \times n^2$, and it is generally constructed (as LeSage and Pace proposed \cite{lesage2008spatial}) as the Kronecker product of the $W$ matrix with itself.
$$W_K = W \otimes W$$

Network theory and spatial econometrics are intimately connected. Leenders \cite{leenders2002modeling} proposed using SAR models employing an ad-hoc $W$ matrix based on the network relations (in terms of social influences and communication); Farber et al. \cite{farber2010topology} analyzed the relationship between the topology property of networks and the properties of spatial models, performing some simulation tests. Mansky's paper \cite{manski1993identification} is seminal, as it laid the foundation for analyzing the exogenous, endogenous and correlated effects that researchers encounter both in network and econometric theory. Lee, Liu and Lin \cite{lee2010specification}, following Mansky's work, proposed a specification and estimation of network econometric models, in the presence of exogenous, endogenous and correlated effects. Furthermore, the discussion on the specification and estimation of the network econometric models has become quite widespread as of late \cite{bramoulle2009identification,chandrasekhar2011econometrics}. To control for network effects, here we use a similar approach and replace the spatial weight the network of trade weight matrix.

\noindent
The specification of the $W$ matrix becomes the following:
\begin{align*}
W: w_{i,j} = & 1 \mbox{ if \textit{i} has a significant trade relationship } \\
& \mbox{with \textit{j} } (p <0.1), w_{i,j}=0 \mbox{ otherwise.}
\end{align*}

\subsection{Model specification}
Le Sage and Pace \cite{lesage2008spatial} proposed to revise spatial specification for OD relationships based on two classes of models: \emph{Spatial autoregressive models} (SAR) and \emph{Spatial Durbin / Spatial error models} (SDM/SEM). The former is motivated by a time relationship describing a diffusion process over space; the latter by the omitted variables argument. We can think of the omitted variable as an unobserved tendency of trade due to network interdependencies.

The econometric representation of the models can be illustrated as:\\
SAR:
\begin{equation}
y = \rho \textbf{W} y + x \beta_1 + \epsilon
\end{equation}
SDM:
\begin{equation}
y = \rho \textbf{W} y + x \beta_1 + \textbf{W} x \beta_2 + \epsilon
\end{equation}
which becomes the SEM model in the event that included and excluded variables are not correlated (Common factor test can be performed \cite{lesage2008spatial}).\\
SEM:
\begin{equation}
y = x \beta_1 + \lambda \textbf{W}\epsilon + \epsilon
\end{equation}

\subsection{Estimation methods}
The concentrated maximum likelihood estimated proposed by Anselin \cite{anselin1988spatial} and revised by LeSage and Pace \cite{lesage2008spatial} generally fits the spatial gravity model of trade. This methods is based on the computation of the log-determinant, which is quite time-consuming. Fortunately, new technical results regarding the calculation of the log determinant have recently come to light, which facilitates estimations \cite{pace2004chebyshev,barry1999monte,smirnov2001fast}. One caveat for this method is that for cases in which a large number of zero flows exists, the method is not appropriate. Maximum likelihood (ML) estimation requires that the dependent variable vector follow a normal distribution, or that it is able to be suitably transformed to achieve normality. A partial solution is generally the logarithmic transformation. Another solution is to consider trade as count data using the Poisson model \cite{lesage2007knowledge}.

\subsection{Data, application and econometric results}
We conduct a cross-section on two different years, 1970 and 2000, on a sample of 146 world countries.
For each year we test a cross section using three different dependent variables: $(i)$ total trade; $(ii)$ differentiated goods; $(iii)$ non-differentiated goods. This allows us to test the Rauch theory, in addition to commenting on the relations between trade of differentiated goods and migration.

The gravity model is a benchmark in international trade and seems to be the best choice for our aims. For the gravity model the trade flow depends on the economic dimension of both the origin and the destination country, and on the distance between the couple. Following this approach, we use the gross domestic product (per capita) and the population as control variables. Furthermore, we control for contiguity, for free trade agreements and for common language and currency, much like what is done in many empirical applications regarding the gravity model of trade. We found all of these variables within the CEPII dataset \cite{mayer2011notes,head2010erosion}. However, the control variable we are most interested in are the migrants stocks: we collected this data from the World Bank dataset \cite{ozden2011earth}.

Appending the spatial autoregressive components in the econometric formula of the gravity model seems to be fundamental in order to grasp the potential contributions of the networks of trade and migration. The concentrated maximum likelihood estimation is the most frequently used estimation method in spatial econometrics because it overcomes issues related to the intrinsic endogeneity in the spatial models. However, the ML estimates require normal distribution, which rarely happens when zero flows are present. We resolve this issue by performing a logarithmic transformation of the dependent and the control variables.

We did a likelihood ratio (LR) and common factor test to choose from different specifications of spatial econometric models, and the results of these tests point in favor of the Durbin model with both the autoregressive component in the dependent variable (trade) and in the control variables (including stock of migration).

In the previous section, we computed four $146\times 146$ matrices for two distinct time periods (the years 1970 and 2000), representing the $(i)$ network of country to country migration; $(ii)$ network of country to country total trades; $(iii)$ network of country to country differentiated goods; $(iv)$ network of country to country non-differentiated goods. We use these matrices as weight matrices in the spatial econometric model, which can be basically represented in this fashion:
\begin{small}
\begin{align*}
t_{00} &= \rho\boldsymbol{W_{t00}}t_{00} + x_{00}\beta_1 + \beta_2 m_{00} + \gamma\boldsymbol{W_{t00}}m_{00}+ \epsilon\\
dt_{00} &= \rho\boldsymbol{W_{tf00}}dt_{00} + x_{00}\beta_1 + \beta_2 m_{00} +\gamma\boldsymbol{W_{tf00}}m_{00}  + \epsilon\\
ndt_{00} &= \rho\boldsymbol{W_{tn00}}ndt_{00} + x_{00}\beta_1 + \beta_2 m_{00} +\gamma\boldsymbol{W_{tn00}}m_{00}  + \epsilon\\
t_{70} &= \rho\boldsymbol{W_{t70}}t_{70} + x_{70}\beta_1 + \beta_2 m_{70} +  \gamma\boldsymbol{W_{t70}}m_{70} + \epsilon\\
dt_{70} &= \rho\boldsymbol{W_{tf70}}dt_{70} + x_{70}\beta_1 + \beta_2 m_{70} +\gamma\boldsymbol{W_{tf70}}m_{70}  + \epsilon\\
ndt_{70} &= \rho\boldsymbol{W_{tn70}}ndt_{70} + x_{70}\beta_1 + \beta_2 m_{70} + \gamma\boldsymbol{W_{tn70}}m_{70}  + \epsilon
\end{align*}
\end{small}
where with $tr$ we represent the total trade between pair of countries, $dtr$ the amount for the differentiated goods and $ndtr$ the non-differentiated goods. 1970 and 2000 remain the time periods. $\rho$ is the coefficient of the lagged trade term, $\beta_1$ the coefficients of the benchmark control variables like \textit{GDP} and population, $\beta_2$ is the migration coefficient and $\gamma$ is the lagged migration coefficient. Moreover, $m$ stay for stock of migration. To conclude, $W_{t}$ is the $n^2 \times n^2$ network of total trade Kronecker matrix, $W_{tf}$ is the $n^2 \times n^2$ network of differentiated trade, $W_{tn}$ is the $n^2 \times n^2$ network of non-differentiated trade and $W_{m}$ is the $n^2 \times n^2$ network of migration.

\begin{table}[!htb]
  \centering
  \caption{Baseline gravity model for Trade - Year: 2000, standard errors in parentheses}
    \begin{tabular}{r|rrrrr}
	\hline
    \rule{0pt}{2ex} & \textbf{Total} & \textbf{Diff.} & \textbf{Homog.} & \textbf{Overlap.} & \textbf{Non-Overlap}\\
    \hline
    \rule{0pt}{3ex} intercept & -7.037*** & -8.869*** & -7.903*** & 5.090*** &5.343*** \\
    & (0.493) & (0.547) & (0.551) & (0.394) & (0.414)\\
		distw & -0.673*** & -0.575*** & -0.570*** & -0.333*** & -0.272*** \\
    & (0.037) & (0.041) & (0.040) & (0.035) & (0.037)\\
    gdpcap\_o & 0.343***  & 0.309*** & 0.353*** & 0.144*** & 0.009\\
    & (0.015) & (0.027) & (0.017) & (0.016) & (0.015)\\
    gdpcap\_d & 0.538*** & 0.606*** & 0.453*** & 0.306*** &0.253*** \\
    & (0.014) & (0.016) & (0.016) & (0.012)& (0.014)\\
    pop\_o & 0.714*** & 0.610*** & 0.783*** & 0.283*** & 0.133*** \\
    & (0.022) & (0.024) & (0.024) &  (0.013) & (0.018)\\
		pop\_d & 1.015*** & 1.146*** & 0.885*** & 0.268*** & 0.260***\\
    & (0.023) & (0.026) & (0.026) & (0.012) & (0.015)\\
    migrat & 0.264*** & 0.238*** & 0.224*** & 0.396*** & 0.322*** \\
    & (0.010) & (0.011) & (0.010) & (0.009) & (0.010)\\
	& & & & &\\
    Adj. $R^2$ & 0.499 & 0.501 & 0.464 & 0.352 & 0.372\\
    \hline
    \end{tabular}
  \label{tab:baseline2000}
\end{table}

\begin{table}[!ht]
  \centering
  \caption{Durbin Model for Trade - Year: 2000, standard errors in parentheses}
    \begin{tabular}{r|rrrrr}
	\hline
    \rule{0pt}{2ex} & \textbf{Total} & \textbf{Diff.} & \textbf{Homog.} & \textbf{Overlap.} & \textbf{Non-Overlap}\\
    \hline
    \rule{0pt}{3ex} intercept & -14.793*** & -10.904*** & -14.275*** & - 10.260*** & -8.672***\\
    & (0.792) & (0.808) & (0.845) & (0.745) & (0.683) \\
	distw & -0.364*** & -0.375*** & -0.373*** & -0.098& -0.220***\\
    & (0.061) & (0.071) & (0.074) & (0.087) & (0.080)\\
    gdpcap\_o & 0.170***  & 0.071** & 0.169*** & 0.315***& 0.126***\\
    & (0.030) & (0.027) & (0.027) & (0.027)& (0.024)\\
    gdpcap\_d & 0.294*** & 0.278*** & 0.256*** & 0.531*** & 0.514***\\
    & (0.026) & (0.023) & (0.024) & (0.026)& (0.025)\\
    pop\_o & 0.713*** & 0.402*** & 0.675*** & 0.367*** & 0.208***\\
    & (0.055) & (0.049) & (0.050) & (0.026)& (0.043)\\
	pop\_d & 0.369*** & 0.326*** & 0.319*** & 0.795*** & 0.892***\\
    & (0.025) & (0.023) & (0.023) & (0.047)& (0.024)\\
    migrat & 0.406*** & 0.391*** & 0.371*** & 0.690*** & 0.593***\\
    & (0.023) & (0.021) & (0.022) & (0.025)& (0.024)\\
    $\gamma$ & 0.009*** & 0.005* & 0.002 & 0.003 & 0.008**\\
    & (0.003) & (0.002) & (0.002) & (0.005)&  (0.004)\\
 	& & & & &\\
    $\rho$  & -0.005 & -0.006 & -0.005 & -0.003 & -0.0001\\
    LR-test & 17.377 & 12.666 & 15.433 & 2.035 & 0.003 \\
    p-value & 3E-05 & 0.0003 & 0.855 & 0.153 & 0.959 \\
    \hline
    \end{tabular}
  \label{tab:durbin2000}
\end{table}

\begin{table}[!ht]
  \centering
  \caption{Durbin Model for Trade - Year: 1970, standard errors in parentheses}
    \begin{tabular}{r|rrrrr}
	\hline
    \rule{0pt}{2ex} & \textbf{Total} & \textbf{Diff.} & \textbf{Homog.} & \textbf{Overlap.} & \textbf{Non-Overlap}\\
    \hline
    \rule{0pt}{3ex} intercept & -5.100*** & -2.671*** & -5.168*** & 0.711 & -0.484\\
    & (0.560) & (0.437) & (0.532)& (0.433) & (0.386) \\
		distw & -0.129** & -0.105* & -0.135* & -0.159** &-0.190***  \\
    & (0.057) & (0.044) & (0.054)& (0.053)& (0.048) \\
    gdpcap\_o & 0.113***  & 0.056*** & 0.095***& 0.154*** & 0.113***  \\
    & (0.015) & (0.012) & (0.014)& (0.019) &(0.017)  \\
    gdpcap\_d & 0.171*** & 0.136*** & 0.155*** & 0.110*** & 0.115***  \\
    & (0.016) & (0.013) & (0.016)& (0.016)& (0.014)  \\
    pop\_o & 0.301*** & 0.111*** & 0.293*** & 0.232*** & 0.176***  \\
    & (0.024) & (0.019) & (0.024) & (0.019) & (0.018) \\
		pop\_d & 0.217*** & 0.158*** & 0.195*** & 0.078** & 0.180*** \\
    & (0.016) & (0.013) & (0.017)& (0.027) &  (0.024)  \\
    migrat & 0.389*** & 0.346*** & 0.352*** & 0.624*** & 0.520*** \\
    & (0.017) & (0.013) & (0.016) & (0.021) & (0.019) \\
    lag.migrat & 0.011*** & 0.003 & 0.010*** & 0.011 & 0.014***  \\
    & (0.002) & (0.002) & (0.002) & (0.003) & (0.003)  \\
	& & & & &\\
    $\rho$   & -0.011 & -0.005 & -0.009 & -0.002 & - 0.004  \\
    LR-test & 54.684 & 7.253 & 33.671 & 1.387 & 9.574  \\
    p-value & 1.4E-09 & 0.007 & 7E-05 & 0.239 & 0.002  \\
    \hline
    \end{tabular}
  \label{tab:durbin1970}
\end{table}

We also perform the baseline gravity model using a new dependent variable representing a novel version of differentiated products that we call \emph{overlapping} (see Section \ref{links_corr}).

We computed these models in $R$ with the $spdep$ package. Results are shown in Tables \ref{tab:baseline2000}, \ref{tab:durbin1970} and \ref{tab:durbin2000}. The most important results can be summarized as follows:
\begin{itemize}
	\item in all models the positive effects on trade of the benchmark variables is confirmed (positive effect for $gdpcap$ and $pop$, negative for distance);
	\item the stock of migrants is highly significant both in 1970 and in 2000, for total trade as well as differentiated and homogeneous goods; however, using our classification of differentiated and homogeneous goods, we can see a more substantial effect of migration on the differentiated ones (see Table \ref{tab:baseline2000});
	\item the lagged stock of migration ($\gamma$) also has a positive impact of trade. This means that, if country $i$ trades with country $j$, and the country $j$ has an high stock of migrants, the trade from $i$ to a third country $k$ will also increase;
	\item using the Rauch classification, this phenomenon is stronger for non-differentiated goods in 1970, while for differentiated goods in 2000. Using our classification, this phenomenon is still not strongly significative for differentiated goods (see Table \ref{tab:durbin2000});
	\item $\rho$ has negative sign for all the models used in Rauch classification. It means that, if a country $i$ imports from country $j$, and the country $j$ has a high level of trade, imports of $i$ from a third country $k$ will decrease (substitution effect). Using the Rauch classification, these coefficients are still negative, altough they are not significative;
	\item the impact of the stock of migrants increases (the coefficient goes from 0.322 to 0.396 in 2000) if we use the new classification. The results confirm our improvements on defining the set of differentiated goods.
\end{itemize}
All in all, the control for network interdependencies is always significant in our analysis. This means that the baseline OD model does not account for network effects, which play a relevant role in shaping the world trade web.



\FloatBarrier

\section{Conclusions}\label{concl}

Sets of interdependent phenomena on a global scale are increasingly analyzed as complex networks. International trade, human mobility, communication and transportation infrastructures are just a few examples. Only recently have new methodologies been developed to analyze the dynamics of intertwined networks, including cascading failures and the transmission of shocks across multiple and heterogeneous network structures. In this paper we contribute to this emerging field of research by analyzing the relationship between the network of human migration and the world trade web. To make different networks comparable, we apply a hypergeometric benchmark to identify significant cross-country relationships. We find that the presence of a stock of migrants in a foreign country has a positive impact on trade. This is particularly true for a subset of products that we identify in Section \ref{network}. However, when we move to test a gravity model of trade controlling for network interdependencies, we do not find any significant difference between the trade of differentiated and homogeneous goods. Since multiple communities of migrants are simultaneously present in developed countries, when we move to consider competition among them and network substitution effects we do not find any particular information advantage for differentiated goods. However, our results in this paper must be considered as preliminary. Spatial econometrics techniques have a huge potential to include network effects in traditional gravity models. In future work we plan to accommodate multiple and different network effects in this setting. Moreover, since the traditional distinction between differentiated and homogeneous goods does not seem to be particularly informative, we plan to investigate the changing role that migrants play in favoring the trade of specific commodities.

\ifCLASSOPTIONcompsoc
  \section*{Acknowledgments}
\else
  \section*{Acknowledgment}
\fi

The authors acknowledge funding from the National Research Program of Italy (PNR), the CRISIS Lab project and the FIRB project ViWaN (Virtual Water Networks). M.R. acknowledges funding from the Multiplex FP7 project (Foundational Research on MULTIlevel comPLEX networks and systems).

\ifCLASSOPTIONcaptionsoff
  \newpage
\fi

\bibliographystyle{IEEEtran}
\bibliography{myrefs}

\end{document}